\documentclass[10pt]{book} 
\usepackage{fix-cm}
\usepackage{amssymb}
\usepackage{amsmath}
\usepackage{graphicx}
\usepackage{subfigure}
\usepackage{makeidx}
\usepackage{multicol}
\usepackage{color}
\usepackage{stmaryrd}
\usepackage{amsmath}%
\usepackage{pgf,tikz}
\usetikzlibrary{arrows}
\usepackage[utf8]{inputenc}
\usepackage{amsthm}

\usepackage{mathtools}
\DeclarePairedDelimiter\ceil{\lceil}{\rceil}

\makeindex

\newtheorem{definition}{Definition}
\newtheorem{theorem}{Theorem}

\DeclareMathOperator*{\argmin}{argmin}
\DeclareMathOperator*{\argmax}{argmax}

\newcommand{\f}{\mathbb{F}_{q}^{n}}
\newcommand{\oo}{\preceq }
\newcommand{\pe}{\varpi}

\newcommand{\Sp}{\textsf{span}}
\newcommand{\calx}{\mathcal{X}}
\newcommand{\cala}{\mathcal{A}}
\newcommand{\calb}{\mathcal{B}}
\newcommand{\calc}{\mathcal{C}}

\newcommand{\cale}{\mathbb{E}}
\newcommand{\calf}{\mathcal{F}}
\newcommand{\calp}{\mathcal{P}}

\newcommand{\cali}{\mathcal{I}}

\newcommand{\caln}{\mathcal{N}}
\newcommand{\su}{\mathrm{supp}}

\newcommand{\il}{\left\langle   }
\newcommand{\ir}{    \right\rangle }

\newcommand{\N}{\mathbb{N}}
\newcommand{\cf}{\mathfrak{c}}

\newcommand{\bu}{\mathbf{u}}
\newcommand{\bv}{\mathbf{v}}

\newcommand{\bx}{\mathbf{x}}
\newcommand{\by}{\mathbf{y}}
\newcommand{\bz}{\mathbf{z}}
\newcommand{\bc}{\mathbf{c}}
\newcommand{\be}{\mathbf{e}}
\newcommand{\bff}{\mathbf{f}}

\begin{document}


\author{Marcelo Firer}
%

\mainmatter

\chapter{Alternative Metrics}\label{AlternativeMetrics}
This is a chapter for "Concise Encyclopedia of Coding Theory" to be published by CRC Press, authored by Marcelo Firer, Imecc - Unicamp.

\section{Introduction}\label{sec:introduction}

The main scope of this chapter is metrics defined for coding and decoding purposes, mainly for block codes. Despite the fact that metrics are nearly ubiquitous in coding theory, it is very common to find them as a general background, with the role of metric invariants not always clearly stated. As a simple but eloquent example, the role of the minimum distance and the packing radius of a code are commonly interchanged. While the minimum distance $d(\calc )$ of a code $\calc $ ensures that any error of size at most $d(\calc  )-1$ can be detected, the packing radius $R(\calc  )$ ensures that any error of size at most $R(\calc  )$ can be corrected. These statements are true for any metric (considering additive errors) and the interchange between these two figures of merits follows from the fact that, in the most relevant metric in coding theory, the Hamming metric, the packing radius is determined by the minimum distance by the famous formula $R(\calc  )=\lfloor \frac{d(\calc  )-1}{2}\rfloor$. This is not the case for general metrics, where not only one of the invariants may not be determined by the other, but determining the packing radius may be an intractable work, even in the case of a $1$-dimensional code (see, for example, \cite{DOLIVEIRA}). As a second example of such a situation is the description of equivalence of codes, found in nearly every textbook on the subject, that seldom mentions the fact that differing by a combination of a permutation of the coordinates and the product by an invertible diagonal matrix means that the two considered codes are image one of the other by a linear isometry of the Hamming space. What is natural to the Hamming metric may become a ``sloppy'' approach when considering different metrics. 

For this reason, besides properly defining a (non-exhaustive) list of metrics that are alternative to the Hamming metric, we will focus our attention on the \emph{structural} aspects of the metric space $(\mathcal{X},d)$. By structural, in this context, we mean aspects  that are relevant for coding theory: the relation between the minimum distance and the packing radius, the weight (or distance) enumerator of the ambient space (or codes), the existence of a MacWilliams type identity, useful description of the group of (linear) isometries of $(\mathcal{X},d)$ and, of course, bounds for the existence of codes with given parameters. What will be mainly missing is the search for codes attaining or approaching the bounds and particular interesting decoding algorithms, among other things, mainly due to space restrictions.

\bigskip

The most important metric in coding theory is the Hamming metric $d_H$. Being an \emph{alternative metric} means, in the context of coding theory, being an alternative to the Hamming metric (or to the Lee metric). The prominence of the Hamming metric is due mainly to the fact that it is \textbf{matched} to the binary symmetric channel, in the sense that (considering codewords to be equiprobable)
\[
\argmin \{d_H(\bx ,\bc ),\bc\in \calc   \} = \argmax \{P_{BSC}(\bc |\bx );\bc\in \calc  \},
\]
for every code $\calc\subseteq\mathbb{F}_2^n$ and every $\bx\in \mathbb{F}_2^n$, where $P_{BSC}(\by |\bx ):= \varrho^{d_H(\bx ,\by )}(1-\varrho)^{n-d_H(\bx ,\by )}$ is the probabilistic model of the channel; see Section \ref{intro:sec:decoding}. 

Since errors are always described in a probabilistic model, \emph{being matched} to a channel is a crucial property of a metric. Despite the fact that any metric can be matched to a channel (the reverse affirmation is not true - see \cite{judy}), only part of the alternative metrics arose matched to a specific channel. Others are considered to be ``suitable'' to correct some types of errors that are characteristic of the channel, despite the fact that the metric and the channel are not matched. Finally,  some of the metrics were studied for their intrinsic interest, shedding some light on the role of metric invariants and  waiting on a theoretical shelf to acquire more practical application. This recently happened to the rank metric, introduced by E. M. Gabidulin in 1985 \cite{gabidulin1985rank} which became outstandingly relevant after the burst in research of network coding; see Chapter \ref{cha:Kschischang}.

\bigskip

All the distance functions surveyed in this chapter actually determine a metric, in the sense they satisfy the symmetry and positivity axioms (what entitles them to be called a \textbf{distance}, according to the terminology used in \cite[Chapter 1]{deza2009encyclopedia}) and also the triangular inequality. At this point we should remark that the triangle inequality, many times the unique property that is not obvious, is a somehow superfluous property in the context of coding theory. Given two distances (or metrics) $d_1$ and $d_2$ defined on $\f$, they should be considered equivalent if they determine the same minimum distance decoding for every possible code $\calc$ and every possible received message $\bx$, in the sense that $\argmin \{d_1(\bx ,\bc ),\bc\in \calc   \}=\argmin \{d_2(\bx ,\bc ),\bc\in \calc   \}  $. Since $\f$ is a finite space, given a distance $d$ it is  always possible to find a metric $d^\prime$ that is equivalent to $d$. For more details on this and related subjects see \cite{oliveira2018}. 

\bigskip
Nearly every metric to be presented in this chapter is a generalization of the Hamming or the Lee metric. Each generalization preserves some relevant properties, while it may fail to preserve  others. In order to understand the way they depart from the Hamming metric, we should list and name some of these properties.

\begin{enumerate}
	\item[WD]\label{WD} \underline{Weight defined}:\index{weight ! metric defined by a} It is \textbf{defined by a weight}: $d(\bx ,\by )=\pe(\bx  -\by )$. This means that the metric is \textbf{invariant by translation}, in the sense that $d(\bx +\bz ,\by +\bz )=d(\bx ,\by ),\forall \bx,\by,\bz\in\mathbb{F}_q^n$. The invariance by translation property makes the metric suitable for decoding linear codes, allowing, for instance, syndrome decoding.
	\item[AP]\label{AP} \underline{Additive Property}:\index{metric ! additive} 
	\index{additive  metric}
	It is  \textbf{additive}, in the sense the weight $\pe$ may be considered to be defined on the alphabet $\mathbb{F}_q$ and $\pe(\bx  )=\sum\pe(x_i)$, for $\bx =x_1x_2\cdots x_n\in\mathbb{F}_q^n$. The additive property is a condition for a metric to be matched to a memoryless channel. 
	\item[RS]\label{RS} \underline{Respect to support}:\index{metric ! which respects support} This means that whenever $\su (\bx  )\subseteq \su (\by )$, we have that $\pe (\bx  )\leq \pe(\by )$. Respecting the support is an essential condition for a metric being suitable for decoding over a channel with additive noise.
	
\end{enumerate}

Nearly all the alternative metrics presented in this chapter are considered in the context of block coding over an alphabet, and they may be essentially distinguished in the way they differ from the Hamming metric in the alphabet, in the spreading over the coordinates or in both ways. They can be classified into some  different types. There are metrics that look in the way errors spread over the bits, ignoring any additional structure on the alphabet. This is the case of the \emph{metrics generated by subspaces} and the \emph{poset metrics}, introduced in Sections \ref{sec:subspacemetrics} and \ref{sec:poset}, respectively. In Section \ref{sec:Lee} we move into a different direction, considering metrics that ``dig'' into the structure of the alphabet and spreads additively over the bits. These are natural generalizations of the Lee metric. In Section \ref{sec:mixing everything} we consider two families of metrics that move in these two directions simultaneously; they consider different weights on the alphabet but are non-additive. Up to this point, all the metrics are defined by a weight, hence invariant by translations. In the next three sections, we approach metrics that are not weight-defined: the metrics for asymmetric channels (Section \ref{sec:asymmetric}), editing metrics, defined over strings of different length (Section \ref{sec:editing}) and metrics defined on a permutation group (Section \ref{sec:permutation}).

Before we start presenting the alternative metrics, we shall establish some notations and conventions: 
\begin{itemize}
	
	\item Given $\bx=x_1x_2\cdots x_n\in\mathbb{F}_q^n$, its \textbf{support}\index{vector ! support of} \index{support of a vector} 
	is denoted by $\su (\bx ):=\{ i\in [n]\mid x_i\neq 0\}$, where $[n]:=\{1,2,\ldots ,n\} $.
	\item $\be_i$ is the vector with $\su (\be_i)=\{i \}$ and the $i$-th coordinate equals $1$. We shall refer to $\beta =\{\be_1,\be_2,\ldots ,\be_n\}$ as the \textbf{standard basis} of $\f$. 
	\item Since we are considering many different metrics, we will denote each \textbf{distance function} by a subscript: the Hamming metric is denoted by $d_H$, the Lee metric by $d_L$ and so on. The subscript is used also in the notation for the \textbf{minimum distance} of a code $\calc$: $d_H(\calc),d_L(\calc)$, etc.
	\item Given a metric $d$ on $\f$, we denote by $B^n_d(r,\bx)=\{\by\in\f \mid d(\bx ,\by)\leq r \}$ and $S^n_d(r,\bx)=\{\by\in\f \mid d(\bx ,\by)=r \}$ the metric \textbf{ball} and \textbf{sphere}, with cardinality $b^n_d(r)$ and $s^n_d(r)$ respectively, the center $\bx$ being immaterial if the metric is  weight-defined. 
	\item The \textbf{packing radius}\index{code !packing radius} $R_d(\calc )$ of a code $\calc$ contained in a space (for instance $\f$) endowed with a metric $d$ is defined by $R_d(\calc )=\max \{r\mid B_d(r,\bx)\cap B_d(r,\by)=\emptyset ,\forall\bx ,\by\in\calc, \bx\neq\by  \} $. 
	\item When considering a metric $d$, we shall say that a code is capable of correcting $t$ errors if the packing radius of the code, according to the metric $d$, equals $t$. 
	\item Since we will consider many different metrics, we may explicitly indicate it by the suffix $d_\ast$, writing, for example, $d_\ast$-weight enumerator, $d_\ast$-perfect, $d_\ast$-isometry, $GL(\mathcal{X}^n,d_\ast )$ and so on.  
\end{itemize}

\section{Metrics generated by subspaces}\label{sec:subspacemetrics}\index{metric !subspace}  This is a large family (increasing exponentially with the length $n$) of non-additive metrics. Metrics generated by subspaces, or simply, \textbf{subspace metrics}, were  introduced by Gabidulin and Simonis in 1998 \cite{projectivemetrics}, generalizing some pre-existing families of metrics. 

Given a set $X\subseteq \mathbb{F}_q^n$, we denote by $\Sp (X)$ the linear subspace of $\f$ spanned by $X$. Let $\mathcal{F}=\{ F_1,F_2,\ldots ,F_m \}$ be a family of subsets of $\mathbb{F}_q^n$ which \emph{generates} the ambient space, in the sense that  $\Sp (\bigcup\limits_{i=1}^mF_i)=\f $.  The elements of $\calf$ are called \textbf{basic sets}\index{basic sets}.

\begin{definition}
	{\em The $\mathbf{\mathcal{F}}$\textbf{-weight} $\pe_\calf (\bx )$ of $\mathbf{0}\neq\bx\in\f$ is the minimum size of a set $I\subseteq [n]$ such that $\bx\in \Sp (\bigcup\limits_{i\in I}F_i) $ and $\pe (\mathbf{0})=0$. The $\calf$\textbf{-distance} is $d_\calf (\bx ,\by )=\pe_\calf (\bx -\by )$.} 
\end{definition}

Without lost of generality, since $\Sp \left({\bigcup\limits_{i\in [n]}}F_i\right)=\Sp \left(\bigcup\limits_{i\in [n]}\Sp (F_i)\right)$, we may assume that each $F_i$ is a vector subspace of $\f$; hence $$\pe_\calf (\bx )=\min \{|I|\mid I\subseteq [m], \bx =\sum_{i\in I}\bx^i,\bx^i\in F_i   \}.$$

The Hamming metric is a particular case of subspace metric, which happens when  $m=n$ and  $F_i=\Sp (\be_i)$, for each $i\in [n]$. The well known \textbf{rank-metric} on the space of all $k\times n$ matrices over $\mathbb{F}_q$, which was introduced in \cite{gabidulin1985rank}, can also be viewed as a particular instance of a projective metric (a special type of subspace metric, to be defined in Section \ref{sec:projective}), where $\calf$ is the set of all $k\times n$ matrices of rank $1$. This metric  became nearly ubiquitous in network coding after Silva, Kschischang and Koetter's work \cite{Danilo}. Despite  its importance, the rank metric will not be explored in this chapter. Many of its important features can be found in Chapter \ref{cha:Kschischang}, devoted to network coding, and in Chapter \ref{cha:Gorla}, where rank-metric codes are examined.

\bigskip

As far as this author was able to track, the subspace metrics, in its full generality, are unexplored in the professional literature. However, something is known about some particular cases, namely projective and combinatorial metrics, most of it due to the work of  Gabidulin and co-authors. We present some of these results on particular subclasses in the sequence.

\subsection{Projective metrics}\label{sec:projective}

Projective metrics were introduced in the same paper \cite{projectivemetrics} where Gabidulin and Simonis first defined the subspace metrics. When each $F_i$ is a vector or a one-dimensional vector subspace of $\f$, it may be seen as a point in projective space; so in this case the $\f$-metric is called a \textbf{projective metric}\index{metric ! projective}. In this situation, we can identify each $F_i$ with a non-zero vector $\bff_i\in F_i$. We denote by 
\[
d_\calf (\calc ) = \min\{d_\calf (\bx ,\by )\mid\bx ,\by\in\calc , \bx\neq\by  \}
\]
the $d_\calf$\textbf{-minimum distance} of a code $\calc$. 
For an $[n,k]_q$ linear code $\calc$, the usual Singleton Bound $d_\calf (\calc )\leq n-k+1$ was proved to hold (see \cite{GabidulinVandermond2003}).

We remark that, for a projective metric, since $\{\bff_1,\bff_2,\ldots ,\bff_m\}  $ generates $\f$, we have that $m\geq n$. By considering the linear map $\varphi:\mathbb{F}_q^m\rightarrow \f$ determined by $\varphi (\be_i)=\bff_i$, its kernel $\calp :=\ker (\varphi )\subseteq  \mathbb{F}_q^m$ is an $[m,m-n]_q$ linear code called the \textbf{parent code}\index{parent code} of $\calf$. Given an $[m,m-n]_q$ linear code $\calc$, the columns of a parity check matrix determine a family $\calf_\calc$ of basic sets such that $\calc$ is the parent code of $\calf_\calc$. The parent code helps to compute the $\calf$-weight of a vector. Given $\bx\in\f$, the inverse image $\varphi^{-1}(\bx )$ is a coset of the parent code $\calp$. If $\by\in\varphi^{-1}(\bx )$ is a coset leader (a vector of minimal Hamming weight),  we have that $\pe_\calf (\bx )=\pe_H(\by)$ \cite[Prop. 2]{projectivemetrics}. 

This simple relation between the $\calf$-weight of a vector and the Hamming weight of the parental coset is used to produce a class of Gilbert-Varshamov Bounds for codes with an $\calf$-metric. We let $L_i(\calp )$ be the number of cosets of $\calp$ with Hamming weight $i$. Then, we have that $L_i(\calp)=s_{d_\calf}(i)$ for every $i$ smaller or equal to the packing radius of the parent code $\calp$. This simple remark is the key to prove the following Gilbert-Varshamov Bound.\index{bound ! Gilbert-Varshamov ! for projective metrics}
\index{Gilbert-Varshamov bound for projective metric} 
\begin{theorem}{\cite[Theorem 3.1.]{projectivemetrics}} 
	Let $\calp_m$ be a sequence of $[m,n]_2$ linear codes with $L_i(\calp_m ) \simeq \binom{m}{i} $, where $\simeq$ means asymptotic behavior. Suppose that $\calp_m$ attains the Gilbert-Varshamov Bound and let $\calf_{m}$ be the family of basic  sets determined by $\calp_m$. Then, there exists a code $\calc\subseteq \f$ with $\calf_{m}$-distance $d$ with $M$ elements, where 
	\[
	M\simeq\frac{2^n}{\sum_{i=0}^{d}\binom{m}{i}}.
	\]
\end{theorem}

The parent code of $\calf$ is  used to produce a family of codes attaining the Singleton Bound, a generalization of Reed--Solomon codes, using concatenation of Vandermonde matrices to generate the basic sets (see \cite[Section 3]{GabidulinVandermond2003}) and also to produce a family of $d_\calf$-perfect codes (see \cite{GabidulinSimonisPerfect} for the construction).

\bigskip

The \textbf{phase--rotation}\index{metric ! phase--rotation} metric is a special instance of a projective metric when $m=n+1$, $F_i=\Sp (\be_i) $ for $i\leq n$ and $F_{n+1}=\Sp (\mathbf{1})$, where $\mathbf{1}:=11\cdots 1$.  It is suitable for decoding in a binary channel in which errors are a cascade of a phase-inversion channel (where every bit of the input message is flipped) and a binary symmetric channel. The actual model of the channel depends on the probability of the two types of errors and hence  the term ``suitable'' is not to be confused with the matching property.

It is easy to see that, in this case, $\pe_\calf (\bx )=\min \{\pe_H(\bx ),n+1-\pe_H(\bx )  \}$. In \cite{phase-rotation}, the authors provide constructions of perfect codes relative to the phase-rotation metric, based on the construction of perfect codes with the Hamming metric and also show how to reduce the decision decoding for the phase-rotation metric to decoding for the Hamming metric. 

.

\subsection{Combinatorial metrics}\label{combinatorialmetrics}

The family of \textbf{combinatorial metrics}\index{metric !combinatorial} was introduced in 1973 by  Gabidulin  \cite{gabidulin1973combinatorial}.  
It is a particular instance of subspace metrics that is obtained as follows: we let $\{A_1,A_2,\ldots ,A_m \mid A_i\subseteq [n] \} $ be a \textbf{covering}\index{covering of a set} of $[n]$, in the sense that $[n]=\cup_{i=1}^{m}A_i$ and define $F_i=\{\bx\in\f \mid\su (\bx )\subseteq A_i  \}$, for $i\in [m]$ and $\calf_\cala =\{F_i\mid 1\leq i\leq m \}$. 

We remark that the combinatorial metric can be expressed in a simpler way, as introduced in \cite{gabidulin1973combinatorial}:   $$\pe_\calf (\bx) = \min \{|I|\mid I\subseteq [m], \su (\bx )\subseteq \bigcup\limits_{i\in I} A_i \}.$$
When considering a combinatorial metric, we shall identify a basic set with its support; hence call each $A_i$ also a basic set and denote $\calf =\{A_1,A_2,\ldots ,A_m \}$. We remark that, if $\mathcal{F}=\{ \{i\}\mid i\in[n]\}$, then $d_{\mathcal{F}}$ is the Hamming metric. 


There are many particular instances of the combinatorial metrics that have been used for particular coding necessities, and there is a substantial literature devoted to codes with ``good'' properties according to such particular instances, despite the fact that the actual metric is hardly mentioned. However, since its introduction in 1973 by Gabidulin,  the structural aspects of the metric were left nearly untouched.  

\index{bound !Singleton ! for combinatorial metrics}
\index{Singleton bound for combinatorial metric} 
A Singleton type bound was determined in 1996 \cite{BossertSidorenko} for codes which are not necessarily linear. We present here its linear version: given an $[n,k]_q$ linear code $\calc$ and a covering $\mathcal{F}$, its minimum distance $d_\mathcal{F}(\calc )$ is bounded by
\[
n\frac{d_\mathcal{F}(\calc )-1}{D}\leq \left\lceil   n\frac{d_\mathcal{F}(\calc )-1}{D} \right\rceil \leq n-k,
\]
where $\lceil \cdot \rceil$ is the ceiling function and $D$ is the minimum number of basic sets needed to cover $[n]$. We remark that $D\leq n$ and equality in the bounds holds if and only if $d_\mathcal{F}$ is the Hamming metric, and, in this case, we get the usual Singleton Bound. 

\medskip

Recently, the combinatorial metrics started to be explored in a more systematic way. In \cite{combinatorialJerry} the authors give a necessary and sufficient condition for the existence of a MacWilliams Identity, that is, conditions of $\calf$ to ensure that  the $d_\calf$-weight distribution of a code $\calc$ determines the $d_\calf$-weight distribution of the dual code $\calc^{\bot}$.

\begin{theorem}
	Let $\calf=\{A_1,A_2,\ldots ,A_m\}$ be a covering of $[n]$. Then, the combinatorial metric $d_\calf$ admits a MacWilliams Identity if and only if $|A_i|$ is constant and $A_i\cap A_j=\emptyset$, for all $i\neq j$.
\end{theorem} 

In the same work, there is a description of the group $GL(\f,d_\calf )$ of linear $d_\calf$-isometries. We briefly explian how it is obtained.

First of all, we say that a permutation $\sigma\in S_n$ \textbf{preserves the covering} if $\sigma (A)\in\calf$, for all $A\in\calf$. Each such permutation induces a linear isometry $T_\sigma (\bx ):=x_{\sigma (1)}x_{\sigma (2)}\cdots x_{\sigma (n)}$. The set of all such isometries is a group denoted by $G_\calf$. 

Next, let $\mathcal{F}^i:=\{A\in \mathcal{F} :  i\in A\}$. It defines an equivalence relation on $[n]$: $i \sim_{\mathcal{F}} j $ if and only if $ \mathcal{F}^i=\mathcal{F}^j$. 
Let $s$ be the number of equivalence classes and denote by $ \llbracket     i \rrbracket $ the equivalence class of  $i$. We construct an $s\times m$ matrix $M=(m_{ij})$, the \textbf{incidence matrix} of this equivalence relation: 

\[
m_{ij}=\begin{cases}
1 \text{ if } \llbracket     i \rrbracket \subseteq A_j, \\ 0 \text{ otherwise}. 
\end{cases}
\]

Given an $n\times  n$ matrix $B=(b_{xy})$ with coefficients in $\mathbb{F}_q$, let us consider the $(i,j)$ block (sub-matrix) $B_{ij}=(b_{xy})_{x\in \llbracket     i \rrbracket, y\in \llbracket     j \rrbracket}$. We say that $B$  \textbf{respects} $M$ if, for $\llbracket     i \rrbracket =\llbracket     j \rrbracket$ the block $B_{ij}=B_{ii}=(b_{xy})_{x,y\in \llbracket     i \rrbracket}$ is an invertible matrix and,  for $\llbracket     i \rrbracket \neq \llbracket     j \rrbracket$, then $B_{ij} \neq \mathbf{0}$ ($B_{ij}$ is a non-null matrix) implies that $\su (\bv^j)\subseteq \su (\bv^i)$, where $\bv^k$ is the $k$-th row of $M$. 
Each matrix $B$ respecting $M$ determines a linear $d_\calf$-isometry and the set $K_M$ of all such matrices is also a group.
\index{group ! linear isometries ! combinatorial metric}

\begin{theorem}
	For $q>2$, the group $GL(\f,d_\calf )$ of linear $d_\calf$-isometries is the semi-direct product  $GL(\f,d_\calf )=G_{\mathcal{F}} \ltimes K_M$. For $q=2$, we have the inclusion $G_{\mathcal{F}} \ltimes K_M\subseteq GL(\mathbb{F}_2^n, d_\mathcal{F})  $ and equality may or not hold, depending of $\mathcal{F}$. 
\end{theorem}

There are some special instances of combinatorial metrics that deserve to be mentioned, namely the block metrics, burst metrics and 2-dimensional burst metrics.

\subsubsection{Block metrics}\label{block} This is the case when $\mathcal{F}$ determines a partition of $[n]$, that is, if $A_i\cap A_j=\emptyset$ for any $A_i,A_j\in\mathcal{F}, i\neq j$. In this context, each basic set is usually called a \textbf{block}. In \cite{feng2006linear} we can find many classical questions of coding theory approached with the block-metrics, starting with the Hamming and Singleton Bounds.\index{metric ! block} 
\index{block  metric}
If we denote $n_i=|A_i|$, the Singleton Bound is just $n-k\geq n_1+n_2+\cdots +n_{d-1}$, where $d$ is the $d_\calf$-minimum distance.  The Hamming Bound is provided by simply counting the number of elements in a metric ball. Both the Singleton and the Hamming Bounds are proved in  \cite[Theorem 2.1]{feng2006linear}. We note that, as a particular instance of a combinatorial metric, the MacWilliams Identity holds for a block-metric only if the basic sets (blocks) all have equal size and in this case, Feng et al. provide an explicit expression for the identity:
\index{MacWilliams ! identity ! for block metrics}
\begin{theorem}{\cite[Theorem 4.3]{feng2006linear}}
	Let $\calc\subseteq\f$ be a linear code and $\calf$ be a partition of $[n]$ with $m$  basic sets, each having cardinality $r$. Let $f_\calc(x ,y )=\sum\limits_{\bc\in\calc}x^{\pe_{\calf}(\calc )}y^{m-\pe_{\calf}(\calc )}$ be the weight enumerator of the code $\calc$. Then, 
	\[
	f_{\calc^{\bot}}(x ,y )=\frac{1}{|\calc |}f_\calc(y-x, y + (q^r - 1)x ).
	\]
\end{theorem}
The authors also establish necessary and sufficient conditions over a parity check matrix for a code to be perfect or MDS. Also, algebraic-geometry codes are explored with this metric.

\subsubsection{$b$-burst metric} \index{metric ! $b$-burst} \index{$b$-burst metric} The  $b$-\textbf{burst metric} is a particular case of combinatorial metric where the basic sets are all the sequences of $b$ consecutive elements in $[n]$, that is, $\mathcal{F}_{[b]}=\{[b],[b]+1,[b]+2,\ldots,[b]+(n-b)\}$, with $[b]+i:=\{1+i,2+i,\ldots ,b+i\}$. A variation will consider all the $n$  cyclic (modulus $n$) sequences of $b$ consecutive elements. It was first introduced in 1970 \cite{burst-metric}. It is suitable for predicting error-correcting capabilities of errors which occur in bursts of length at most $b$. 

\begin{theorem}{\cite[Theorems 1,2 and 4]{burst-metric}}
	Let $\calc$ be a linear code with minimal burst distance $d_\calf (\calc )$, where $\calf = \calf_{[b]}$. Then, it corrects:
	\begin{enumerate}
		\item Any pattern of $\lfloor \frac{d_\calf (\calc )-1}{2} \rfloor$  error bursts of length $b$;
		\item Any pattern of $d_\calf (\calc )-1$ erasure bursts of length $b$;
		\item Any pattern of $m_1$  error bursts and $m_2$  erasure bursts of length $e_1,e_2,\ldots ,e_{m_2}$, with every erasure  preceded and followed by a burst error of length $b_{1,i}$ and $b_{2,i}$ respectively, provided that $2m_1+m_2<d_\calf (\calc)$ and $b_{1i}+e_i+b_{2i}\leq b  $. 
	\end{enumerate} 
\end{theorem}

As an example, it is possible to prove that a code $\calc$ can correct any burst error of length $b$ if and only if $3\leq\min \{d_{\mathcal{F}_{[b]}}(\bx ,\by)\mid \bx ,\by\in \calc, \bx\neq\by \} $, that is, the error correction capability (the packing radius) is determined by the minimum distance and this happens in a similar way to the Hamming metric.  The  Reiger Bound \cite{Reiger}, for example, can be expressed in terms of the packing radius.

\medskip

A different instance where the $b$-burst metric arises is due to high density storage media. In this storage devises, the density does not permit to read the bits one-by-one, but only on sequences of $b$-bits at a time. This is what is called the $\mathbf{b}$\textbf{-symbol read channel}. 
Recently, Cassuto and Blaum (\cite{cassutob}, for the case $b=2$) and  Yaakobi et al. (\cite{Yaakobib}, for $b>2$), proposed coding and decoding schemes for the $b$-symbol read channel. The decoding scheme is a nearest-neighbor decoding, which consider a metric that is actually the $b$-burst metric. Many bounds on such coding schemes and construction of MDS-codes can be found in the literature (see for example \cite{DING2018}). We should remark that in all these works, the authors consider as undistinguished a particular encoding approach and the metric decoding.

\subsubsection{$b_1\times b_2$-burst metrics}\index{metric ! $b_1\times b_2$-burst}
\index{$b_1\times b_2$ metric}

When considering codewords to be represented as a matrix, a way to couple with burst of size $b_1$ and $b_2$ in the rows and columns respectively,  is to consider as the basic sets the $(b_2+1)\times (b_1+1)$ matrices
\[
T_{r,s}=\left(   
\begin{array}{cccc}
(r,s)&(r,s+1)  &\cdots  &(r,s+b_1)  \\ 
(r+1,s)&(r+1,s+1)  &\cdots  &(r+1,s+b_1) \\ 
\vdots & \vdots &\ddots  & \vdots \\ 
(r+b_2,s)&(r+b_2,s+1)  &\cdots  &(r+b_2,s+b_1)
\end{array} 
\right),
\]
where $r\in [n_1],s\in [n_2]$ and the entries are considered $(\mod n_1,\mod n_2)$. A large account on the subject can be found in survey  \cite{arraycodes} on array codes.

\section{Poset metrics}\label{sec:poset}

The poset metrics, in its full generality, were introduced by Brualdi, Graves and Lawrence in 1995 \cite{Brualdi-poset}, as a generalization of Niederreiter's metric \cite{Niederreiter1992combinatorial}. Few of these metrics are matched to actual relevant channels, but they are relatively well understood in their intrinsic aspects and so becoming an interesting contribution for the understanding of the role of metrics in coding theory. As the subspace metrics presented in Section \ref{sec:subspacemetrics}, it is a family of non-additive metrics, respecting support and defined by a weight (properties {RS} and {WD} in Section \ref{sec:introduction}). 

In what follows we shall introduce the poset metrics as presented in \cite{Brualdi-poset}, then consider some particular cases of posets and  later move on to the many generalizations that arose in recent years. A recent account on poset metrics and its generalizations can be found in \cite{livro}.

\bigskip
Let $P=([n],\oo_P )$ be a partial order (\textbf{poset}\index{poset, partialy ordedered set}, for short) over $[n]$. A subset $I\subseteq [n]$ is called an \textbf{ideal} if $i\in I$ and $j\oo_P i$ implies $j\in I$. For $X\subseteq [n]$, we denote by $\langle X \rangle_P$ the \textbf{ideal generated by} $X$ (the smallest ideal containing it). Given $\bx\in \mathbb{F}_q^n$, the $P$-\textbf{weight} of $\bx$ is $\pe_P(\bx  )= |\langle \su (\bx  ) \rangle_P |$, where $|A |$ is the cardinality of $A$ and $\su (\bx  )$ is the support of the vector $\bx$. The function $\pe_P$ is a weight and the \textbf{poset metric} \index{metric ! poset}\index{poset ! metric} $d_P$ is the weight defined metric $d_P(\bx ,\by )=\pe_P(\bx  -\by )$ 
We should remark that for the particular case of an \textbf{anti-chain}\index{poset ! anti-chain} ($i\oo_P j \iff i=j$) we have that the poset metric coincides with the Hamming metric.  It is worth remarking that  this family is complementary to the family of metrics generated by subspaces (Section \ref{sec:subspacemetrics}), in the sense that the Hamming metric is the only intersection of these two families.

Codes with a poset metric, or simply poset codes, were investigated in many different paths, including conditions (on the poset) to turn a given family of codes into perfect or MDS codes. The fact that there are relatively many codes with such properties is what first attracted the attention to poset metrics. For perfect or MDS codes with poset metrics see \cite{Brualdi-poset, Hyun200437, Kim20085241, Jang200385, DS, Quistorff20072514} and \cite{Hyun20118021}, among others. 

Considering more structural results, one can find in \cite{Panek20084116} a description of the group of linear isometries of the metric space $(\mathbb{F}_q^n,d_P)$ as a semi-direct product of groups. Each element $\sigma$ of the group $Aut(P)$ of poset automorphisms (permutations of $[n]$ that preserve the order) determines a linear isometry $T_\sigma (\bx)=x_{\sigma (1)}x_{\sigma (2)}\cdots x_{\sigma (n)}$. The set of all such isometries is a group, also denoted by $Aut(P)$. We let $M_P$ be the order matrix of $P$, that is, $M_P$ is an $n\times n$ matrix and $m_{ij}=1$ if $i\preceq j$ and $m_{ij}=0$ otherwise. An $n\times n$ matrix $A$ with entries in $\mathbb{F}_q$ such that $a_{ii}\neq 0$ and $\su (A)\subseteq\su (M_P)$ determines a linear isometry and the family of all such maps is a group, denoted by $G_P$.
\index{group !  linear isometries ! of a poset metric}
\begin{theorem}{\cite[Corollary 1.3]{Panek20084116}}
	The group $GL(\f,d_P)$ of linear isometries of $(\f ,d_P)$ is the semi-direct product $GL(\f,d_P)=Aut(P)\ltimes G_P$.
\end{theorem} 
MacWilliams-type identities are explored  in \cite{choi2012macwilliams}, revealing a deep understanding about the most important duality result in coding theory. First of all, it was noticed since \cite{Jang200385} that, when looking on the dual code $\calc^{\bot}$, we are actually not interested in the $d_P$-weight distribution but in its $d_{P^\bot}$-weight distribution, determined by the \textbf{opposite poset}\index{poset !opposite} $P^\bot$ defined by $i\oo_{P^\bot}j\iff j\oo_Pi$. The innovation proposed in \cite{choi2012macwilliams} is to note that the metric invariant is not necessarily the weight distribution, but rather a distribution of elements in  classes determined by a suitable equivalence relation on ideals of $P$. The authors identify three very basic and different equivalence relations of ideals. Given two poset ideals $I,J\subseteq [n]$ we establish the following equivalence relations. 

\begin{itemize}
	\item[\textbf{($E_C$)}]: $(I,J)\in E_C$ if they have the same cardinality. 
	\item[\textbf{($E_S$)}]: $(I,J)\in E_S$ if they are isomorphic as sub-posets.
	\item[\textbf{($E_H$)}]: Given a subgroup $H\subseteq Aut(P)$, $(I,J)\in E_H$ if there is $\sigma\in H$ such that $\sigma (I)=J$.
\end{itemize}

It is easy to see that $E_H\subseteq E_S \subseteq E_C$. 

Given an equivalence relation $E$ over the set $\cali (P)$ of ideals of $P$, for each coset $\overline{I}\in\cali (P)/ E$, the $\overline{I}$-sphere $S^P_{\overline{I},E}$ centered at $\textbf{0}$ with respect to $E$ is 

\[
S^P_{\overline{I},E}:=\{\bx\in\f\mid (\il \su (\bx )\ir_P,I)\in E  \}.
\]

If $E$ is an equivalence relation on $\cali (P)$, we denote by $E^\bot$ the equivalence relation on $\cali (P^\bot )$, where $E$ may be either $E_C,E_S$ or $E_H$. Given an ideal $I\in\cali (P)$, we denote its complement by $I^c=[n]\setminus I$. We say that $E^\bot$ is the \textbf{dual relation} of $E$.\index{poset ! dual relation} 

We define $A_{\overline{I},E}(\calc ):= |S^P_{\overline{I},E}\cap\calc |$ and call $W_{P,E}(\calc ):=[A_{\overline{I},E}(\calc ) ]_{\overline{I}\in \cali (P)/E}$ the \textbf{spectrum} of $\calc$ with respect to $E$.\index{code ! spectrum} We remark that, for $E=E_C$, the spectrum of $\calc$ consists of the coefficients of the weight enumerator polynomial of $\calc$. 

\begin{definition}
	{\em Let $P$ be a poset, $E$ an equivalence relation on $\cali (P)$ and suppose that $E^\bot$ is the dual relation on $\cali (P^\bot )$. The relation $E$ is a \textbf{MacWilliams equivalence relation}\index{MacWilliams ! equivalence relation} if, given linear codes $\calc_1$ and $\calc_2$, $W_{P,E}(\calc_1 )=W_{P,E}(\calc_2 )$ implies $W_{P^\bot,E^\bot}(\calc_1^\bot )=W_{P^\bot,E^\bot}(\calc_2^\bot )$.}
\end{definition} 

For the case $E=E_C$, to be a MacWilliams equivalence implies the existence of a MacWilliams Identity in the usual sense. In general, for $E$ being a MacWilliams equivalence relation depends on the choice of $P$:
\index{MacWilliams ! identity ! for poset metrics}
\begin{theorem}{\cite[Theorem 3]{choi2012macwilliams}}
	\begin{enumerate}
		\item $E_H$ is a MacWilliams equivalence relation, for every poset $P$ and every subgroup $H\subseteq Aut(P)$.
		\item $E_S$ is a MacWilliams equivalence relation if and only if $P$ is such that $I\sim J$ implies $I^c\sim J^c$, for every $I,J\in\cali (P)$.
		\item $E_C$ is a MacWilliams equivalence relation if and only if  $P$ is a hierarchical poset.
	\end{enumerate}
\end{theorem} 

As for the last item, a poset $P$ is called \textbf{hierarchical}\index{poset ! hierarchical} if $[n]$ can be partitioned into a disjoint union $[n]=H_1\cup H_2 \cup \cdots \cup H_r$ in such a way that $i\oo_P j$ if and only if $i\in H_{h_i},j\in H_{h_j}$ and $h_i<h_j$. This family of posets includes the minimal poset (an anti-chain) and maximal (a chain,  see definition on page  \pageref{chain}). Hierarchical poset metrics are a true generalization of the Hamming metric, in the sense that many well known results concerning the Hamming metric are actually a characteristic of hierarchical poset metrics.

\begin{theorem}{\cite{hierachical}}
	Let $P=([n],\oo_P)$ be a poset. Then, $P$ is hierarchical if and only if  any of the following  (equivalent) properties holds:
	\begin{enumerate}
		\item $P$ admits a MacWilliams Identity  ($E_C$ is a MacWilliams equivalence relation).
		\item The MacWilliams extension property holds: given two $[n,k]_q$ linear codes  $\calc_1,\calc_2$, any linear isometry $f:\calc_1\rightarrow\calc_2$  can be extended to a linear isometry $F:\f\rightarrow\f$.
		\item The packing radius of a code is a function of the minimum distance.
		\item Linear isometries act transitively on metric spheres: for every $\bx,\by\in\f$, if $\pe_P(\bx )=\pe_P(\by ) $, there is $\sigma\in GL(\f,d_P)$ such that $\sigma (\bx )=\by$.
	\end{enumerate}
\end{theorem} 

All these properties (and some more in \cite{hierachical}) are a direct consequence of the canonical decomposition of codes with hierarchical posets \cite{Felix2012315}, which allows the transformation of problems with hierarchical poset metrics into problems with the well studied Hamming metric.

\bigskip

Another relevant class of poset-metrics is the family of NRT metrics, defined by a poset which is  a disjoint union of chains. A poset $P$ is a \textbf{chain}\label{chain}\index{poset ! chain}, or total order, when (up to relabeling) $1\oo_P 2\oo_P \cdots \oo_P n$. Let $(P,[n])$ be a disjoint union of $m$ chains, each of length $l$; that is, $n=ml$ and the relations in $P$ are $jl+1\oo_P jl+2\oo_P \cdots \oo_P (j+1)l$, for $j=0,1,\ldots ,m-1$. This is the original instance explored by Niederreiter in a sequence of three papers \cite{Niederreiter1987smalldiscrepancy,Niederreiter1992,Niederreiter1992combinatorial} and later,  in 1997, by Rosenbloom and Tsfasman
\cite{Rosenbloom1997}, with an information-theoretic approach. After these three main authors, it is known as the \textbf{NRT metric}\index{poset ! metric ! NRT }. \index{NRT metric}
Many coding-theoretic questions have been investigated with respect to the NRT metric, including MacWilliams Duality \cite{1}, MDS codes \cite{4,Skriganov2001uniform-distributions}, 
structure and decoding of linear codes \cite{NRT,7} and coverings \cite{coveringNRT}. In \cite{BP15} the authors show the connection of NRT metric codes with simple models of information transmission channels and introduce a simple invariant that characterizes the orbits of vectors under the group of linear isometries (the \textbf{shape} of a vector). For the particular case of one single chain ($n=l,m=1$), $d_P$ is an ultra-metric and, given a code $\calc$ with minimum distance $d_{d_P}(\calc)$, it is possible to detect and correct every error of $P$-weight at most $d_{d_P}(\calc )-1$. Despite some strangeness that may be caused by ultra-metrics, this instance is simple and well understood (see \cite{NRT}).

\medskip

An extensive and updated survey of poset metrics in the context of coding theory can be found in \cite{livro}.
In the sequence we will present some recent generalizations of the poset metrics.

\subsection{Poset-block metrics}\label{poset_block}

In 2006, Feng et al. \cite{feng2006linear} started to explore the block metrics, a special instance of the combinatorial metrics, presented in Session \ref{block}.  In 2008, Alves et al. \cite{poset-block} combined the poset and the block structure, giving rise to the so-called poset-block metrics.

We let 
\[
\mathbb{F}_q^N:=V_1\oplus V_2\oplus \cdots \oplus V_n
\]
be a decomposition of $\mathbb{F}_q^N$ as a direct sum of subspaces. We denote $k_i=\dim (V_i)>0$, $\pi =(k_1,k_2, \ldots ,k_n)$ and  we call this decomposition a \textbf{block structure}\index{block structure}.    Being a direct sum we have that each $\bx\in \mathbb{F}_q^N$ has a unique decomposition as $\bx=\bx_1+\bx_2+\cdots +\bx_n$, with $\bx_i\in V_i$. Considering such a decomposition  the \textbf{$\pi$-support} of $\bx$ is defined as $\su_{\pi}(\bx):=\{i\in [n]\mid\bx_i\neq 0\}$. Counting the $\pi$-support gives rise to the \textbf{$\pi$-weight} $\pe_{ \pi}(\bx):=| \su_{\pi}(\bx)|$.

If $P=([n],\preceq_P )$ is a poset over $[n]$, we may combine the block structure $\pi$ and the poset structure $P$ into a single one, by counting the $P$-ideal generated by the $\pi$-support of a vector, obtaining the \textbf{poset-block weight} and the \textbf{poset-block metric}:\index{metric ! poset-block} 

\[
\pe_{(P,\pi )}(\bx):=|\il \su_{\pi}(\bx)\ir_P| \hspace{3pt} \text{ and } \hspace{3pt} d_{(P,\pi )}(\bx,\by):=\pe_{(P,\pi )}(\bx-\by)
\]

In \cite{poset-block}, the authors give a characterization of the group of linear isometries of $(\mathbb{F}_q	^n,d_{(P,\pi )})$. Necessary and sufficient conditions are provided for such a metric to admit a MacWilliams Identity. Perfect and MDS codes with poset-block metrics are investigated in \cite{poset-block} and \cite{Dass2017}. It is worth noting that an interesting feature of this family of metrics is combining together a poset metric, which increases the weight of vectors (hence ``shrinking'' the metric balls), with a block metric, which decreases the weights (hence ``blowing-up'' the balls).

\subsection{Graph metrics}\label{graph}

The Hasse diagram of a poset $P$ can be seen as special kind of directed graph, just thinking of an edge  of the diagram connecting $i,j\in [n]$ to be a directed edge with initial point at $j$ if $i\oo_P j$.  Inasmuch, it is natural to generalize the poset metrics to digraphs. This family of metrics was introduced in 2016 in \cite{Etzion1} and explored in \cite{Etzion}. 

Consider a  finite \textbf{directed graph}\index{directed graph}  (or simply \textbf{digraph}) $G=G(V,E)$ consisting of a finite set
of \textbf{vertices} $V=\{v_1 ,\ldots , v_n\}$ and a set
of \textbf{directed edges} $E\subseteq V\times V$ (parallel edges are not allowed). A \textbf{directed path} is a sequence $v_{i_1},v_{i_2},\ldots ,v_{i_r}$ of vertices where every two consecutive vertices determine a directed edge. A \textbf{cycle} is a  directed path  in which $v_{i_1}=v_{i_r}$. We say that a vertex  $u$ \textbf{dominates} a vertex $v$ if there is a directed path starting at $u$ and passing trough $v$. A set $X\subseteq V$ is called a \textbf{closed set}
if $u\in X$ and $u$ dominates $v\in V$ implies that $v\in X$. The \textbf{closure}
$\langle X\rangle_G $ of a set $X\subseteq V$ is the smallest closed subset containing $X$.

%

We identify $V=\{ v_1 ,\ldots , v_n \}$ with $[n]=\{ 1,2, \ldots ,n\}$ and define the $G$-weight $\pe_G(\bx)$ of $\bx\in\f$ as the number of  vertices in $G$ dominated by the vertices in the support of $\bx$:
\[
\pe_G(\bx):= |\il \su (\bx)\ir_G |.
\]
The $G$-\textbf{distance} between  $\bx, \by \in \f$ is defined by $d_G(\bx,\by):=\pe_G(\by-\bx)$. \index{metric ! graph}
\index{graph metric}

In the case that $G$ is \textbf{acyclic} (that is, contains no cycles), the metric $d_G$ is actually a poset metric: the existence of a non-trivial cycle would contradict the anti-symmetry axiom of a poset ($i\preceq_P j $ and $j\preceq_Pi$ implies $i=j$).  Among the few things that are known about such metrics, there are two canonical forms of a digraph which are able to establish whether two different graphs determine the same metric. Also, similarly to the poset-block case, in \cite{Etzion} there is a reasonable description of the group of linear isometries and some sufficient (not necessary) conditions for a digraph-metric to admit the extension property and the MacWilliams Identity. It is worth to note that the conditions for the validity of the extension property and the MacWilliams Identity do not coincide. As far as the author was able to look, this is the first instance where the hypotheses for the MacWilliams extension property and the  MacWilliams Identity are different.

A different formulation of this digraph metric can be obtained by considering weights on a poset, as defined in \cite{weightedposet}, where perfect codes start to be explored in this context.

\section{Additive generalizations of the Lee metric}\label{sec:Lee}

In this section we present  some  families of metrics defined over an alphabet $\calx$, generally a ring $\mathbb{Z}_q$ (or a field $\mathbb{F}_q$) which generalize the Lee metric. They are all additive metrics, depending on the definition of a weight on the base field (or ring). 

To study different weights on $\mathbb{Z}_m$ (or other algebraic structures) means to describe (and possibly classify) all possible weights, up to an equivalence relation that arises naturally in the context of coding theory:  two weights (or metrics) are considered to be equivalent if they determine the same collection of metric balls (see \cite{DOliveira2018} for a precise definition).
The problem of describing all non-equivalent weights on $\mathbb{Z}_q$ was raised by Gabidulin in \cite{gabidulinsurvey}, and discussed for $m=4$. A general approach to this problem was made in a yet unpublished Ph.D. thesis (see \cite[Theorem 2.3.8]{Max}, in Spanish) where  all possible weights are classified, assuming the alphabet to be just an Abelian group. 

\subsection{Metrics over rings of integers}\label{sec:Gaussian}

Metrics on the quotient rings of Gaussian integers were first introduced by  Huber in \cite{HuberGauss} and they admit many variations and generalizations. They are considered suitable for signal constellations as QAM. We introduce them here in their simplest and most usual setting.

Given $x+iy\in\mathbb{Z}[i]$, we consider the norm $\caln (x+iy)=x^2+y^2$. Let $\alpha=a+bi\in\mathbb{Z}[i]$ be a Gaussian prime, that is, either $a^2+b^2=p$ is a prime and $p\equiv 1\pmod 4$ or $\alpha =p+0i$ with $p\equiv 3\pmod 4$. Given $\beta\in\mathbb{Z}[i]$, there are $q,r\in\mathbb{Z}[i]$ such that $\beta=q\alpha+r$, with $\caln (r)<\caln (\alpha )$. We denote by $\mathbb{Z}[i]/\il\alpha\ir$ the ring of the equivalence classes of $\mathbb{Z}[i]$ modulo the ideal generated by $\alpha$. For $\beta\in\mathbb{Z}[i]/\il\alpha\ir$, let $x+iy$ be a representative of the class $\beta$ with $|x|+|y|$ minimum. The $\alpha$\textbf{-weight} $\pe_\alpha$ and the $\alpha$\textbf{-distance} $d_\alpha$ are defined by $\pe_\alpha (\beta ):=|x|+|y |$ and $d_\alpha (\beta ,\gamma )=\pe_\alpha (\beta -\gamma)$, respectively. 

The $\alpha$-weight and $\alpha$-distance are extended additively to the product $(\mathbb{Z}[i]/\il\alpha\ir)^n$ by setting \[
\pe_\alpha (\mathbf{\beta})=\sum_{i=1}^n\pe_\alpha(\beta_i ) \text{ and } d_\alpha (\mathbf{\beta} ,\mathbf{\gamma} )=\pe_\alpha (\mathbf{\beta} -\mathbf{\gamma} )
,\]
for $\beta=\beta_1\beta_2\cdots \beta_n\in (\mathbb{Z}_\alpha)^n$. The distance $d_\alpha$  is known as the \textbf{Manhheim metric}.\index{metric ! Manhheim}
\index{Manhheim metric}

There are many variations of the Mannheim metric, obtained by considering other rings of integers in a  cyclotomic field. The same kind of construction can be done, for example in the ring of \textbf{Eisenstein-Jacobi integers} (see \cite{huberEJ}). We let $\zeta =(-1+i\sqrt{3})/2$ and consider $\mathbb{Z} [\zeta]:=\{x+y\zeta \mid x,y\in\mathbb{Z}  \}$. We consider a prime $p\equiv 1 \pmod 6$, which can be expressed as $p=a^2+3b^2=\alpha\alpha^\ast$, where $ \alpha =a+b+2b\zeta$ and $\alpha^\ast=a+b+2b\zeta^2$ and let $\mathbb{Z}[\zeta ]/\il\alpha\ir$ be  the quotient  ring. Given $\beta\in\mathbb{Z}[\zeta ]/\il\alpha\ir$, we define $\pe_{ \zeta,\alpha}(\beta )$ to be the minimum of $|x_1|+|x_2|$, where $\beta =x_1\epsilon_1+x_2\epsilon_2$ with $\epsilon_1,\epsilon_2\in\{\pm 1,\pm\zeta,\pm (1+\zeta)  \}$. The definition of a metric on $\left( \mathbb{Z}[\zeta ]/\il\alpha\ir \right)^n$ is made in a similar way, considering the distance on $\mathbb{Z}[\zeta ]/\il\alpha\ir$ determined by the weight and extending it additively to the $n$-fold product of $\mathbb{Z}[\zeta ]/\il\alpha\ir$.

We remark that codes over rings of integers can be considered as codes on a flat torus, a generalization of spherical codes (see, for example, \cite{SueliCosta} and \cite{MartinezGabidulin}).

Most of the results using metrics over rings of integers are constructions of codes that are able to correct a certain number of errors, along with decoding  algorithms (\cite{huberEJ,HuberGauss}). It is also worth noting that these metrics can be obtained as  path-metrics on appropriate circulant graphs, as presented in \cite{MartinezGabidulin}.

\subsection{$l$-dimensional Lee weights}

This is a family of metrics defined by S. Nishimura and T. Hiramatsu in 2008 \cite{NISHIMURA}. It is an interesting construction that has not yet been explored in the literature. We let $\beta =\{\be_1,\be_2,\ldots ,\be_l\}$ be the standard basis of $\mathbb{R}^l$, and we consider the finite field $\mathbb{F}_q$, $ q=p^m$. Let $\phi:\beta \rightarrow \mathbb{F}_q$ be a  map and denote $a_i:=\phi (\be_i)$. Given $\bu=\sum_{i=1}^{l}u_i\be_i\in\mathbb{Z}^l$, we define $\phi (\bu)= \sum_{i=1}^{l}a_iu_i\in \mathbb{F}_q$. We shall assume that the $a_i$'s are chosen in such a way that $\phi$ is surjective, so that given $a\in\mathbb{F}_q$ there is $\bu$ such that $\phi (\bu)=a$. 

The $l$-dimensional $\phi$ weight  $\pe_{l,\phi}(a)$ of $a\in\mathbb{F}_q$ is defined by \[\pe_{l,\phi}(a):=\min  \{ \sum_{i=1}^l|x_i| \mid \bx\in \phi^{-1}(a) \} .\]
The  $l$-\textbf{dimensional Lee weight } of $\bx\in\f$ is defined additively by $\pe_{l,\phi}(\bx) :=\sum_{i=1}^n\pe_{l,\phi}(x_i)$. The weight of the difference $d_{l,\phi}(\bx ,\by)=\pe_{l,\phi}(\bx -\by )$ is the $l$-\textbf{dimensional Lee distance} determined by $\phi$. \index{metric ! $l$-dimensional Lee}
\index{$l$-dimensional Lee  metric}
\medskip

The $l$-dimensional Lee distance is a metric which generalizes some known metrics. This is the case of the {Lee metric}, which is obtained by setting $l=1$ and  $\phi(\be_1)=1$. Also the {Mannheim metric} can be obtained as a $2$-dimensional Lee metric: for $p\equiv 1\pmod 4$, the equation $x^2 \equiv -1\pmod p$ has a solution $x=a\in\mathbb{F}_p$. For $l=2$, we define $\phi(\be_1)=1$ and $\phi (\be_2 )=a$ and one gets the Mannheim metric (Section \ref{sec:Gaussian}) as a particular instance of a $2$-dimensional Lee metric.
\bigskip

In \cite{NISHIMURA} there are constructions of codes correcting any error $\be$ of weight $\pe_{l,\phi}(\be )=1$ (for $q$ odd) and $\pe_{l,\phi}(\be )=2$ (for the case $q=p^m=4n+1, p>5$).

\subsection{Kaushik-Sharma metrics}

A \textbf{Kaushik-Sharma metric}\index{metric ! Kaushik-Sharma} \index{Kaushik-Sharma metric} is an additive metric over $\mathbb{Z}_q$, for an integer $q>1$,  which generalizes  the Lee metric. It is named after M. L. Kaushik, who introduced it in \cite{KAUSHIK} and B. D. Sharma, who studied it in subsequent works. We say that a partition $\calb = \{B_0,B_1,\ldots ,B_{m-1}\}$ of   $\mathbb{Z}_q$ is a \textbf{KS-partition} if it satisfies the following conditions:
\begin{enumerate}
	\item $B_0=\{0 \}$ and, for $i\in\mathbb{Z}_q\setminus \{0\}$, $i\in B_s \iff q-i\in B_s$;
	\item If $i\in B_s,j\in B_t$ and $s<t$ then $\min \{ i,q-i\}<\min \{ j,q-j\}$;
	\item $|B_0|\leq |B_1|\leq \cdots \leq |B_{m-2}|$ and $ \frac{1}{2}|B_{m-2}|\leq |B_{m-1}|$.
\end{enumerate}

Given $j\in\mathbb{Z}_q$, there is a unique $s\leq m-1$ such that $j\in B_s$ and we write $\pe_\calb (j)=s$. The \textbf{KS-weight} and \textbf{KS-distance} determined by the KS-partition $\calb$ are defined, respectively, by
\[
\pe_\calb (\bx)=\sum_{\i=1}^{n}\pe_\calb (x_i) \text{ and } d_{\calb}(\bx ,\by )=\pe_{ \calb}(\bx -\by),
\]
for $\bx ,\by \in \mathbb{Z}_q^n$.	
The KS-distance determines a metric which generalizes the Hamming metric (for the partition $\calb_H$ defined by $B_1=\{1,2, \ldots ,q-1\}$) and the Lee metric (for the partition $\calb_L$ defined by $B_i=\{i,q-i\}$, for every $1\leq i \leq \lfloor 1+ q/2  \rfloor$).

Considering  $\mathbb{Z}_q^n$ as a module over $\mathbb{Z}_q$, we say that $\calc\subseteq\mathbb{Z}_q^n$ is \textbf{linear over} $\mathbb{Z}_q$ if it is a $\mathbb{Z}_q$-submodule of $\mathbb{Z}_q^n$; see Chapter \ref{cha:Dougherty}. Since $\mathbb{Z}_q$ is commutative, it has invariant basis number and we denote by $k$ the rank of $\calc$.
Considering a KS-metric, the main existing bound is a generalization of the Hamming Bound, and it applies for linear codes over $\mathbb{Z}_q$. 
\index{bound ! Hamming ! for Kaushik-Sharma metrics} 

\begin{theorem}{\cite[Theorem 1]{KAUSHIK}}
	Let $\calc\subset\mathbb{Z}_q^n$ be a linear code of length $n$ over $\mathbb{Z}_q$ that corrects all errors $\be$ with $\pe_\calb(\be )\leq r_1$ and all errors $\textbf{f}$ consisting of bursts of length at most $b<n/2$ and  $\pe_{ \calb}(\textbf{f} )\leq r_2$, with $1<r_1<r_2<(m-1)b$. Then,
	\[
	n-k\geq \log_q b^n_{d_{\calb} }(r_1) +\sum_{i=1}^b(n-i+1)(b_{d_{\calb}}^i(r_2)-b_{d_{\calb}}^i(r_1)      ),
	\] 
	where $b_{d_{\calb}}^i(r)$ is the cardinality of a metric ball of radius $r$ (in $\mathbb{Z}_q^i$).
\end{theorem}

In more recent works \cite{sharmaGaur,sharmaGaur2}, there is a refinement of the bounds, by considering errors of limited pattern, that is, by possibly limiting the range of errors in each coordinate: $\pe_{ \calb}(x_i)\leq d$. It is worth noting that a  generalization of the Kaushik-Sharma metric can be done for any finite group (not necessarily cyclic), as presented in \cite{batagel}.

\section{Non-additive metrics digging into the alphabet}\label{sec:mixing everything}

In the previous sections, we considered metrics that either ignore different weights on the alphabet and look for the way it spreads over different positions (as in Sections \ref{sec:subspacemetrics} and \ref{sec:poset}) or metrics which look into the alphabets and spread it additively over a finite product of the alphabet (as in Section \ref{sec:Lee}). 

In this section we consider two families of metrics that dig into the alphabet, i.e. consider different weights on the alphabet, and are not necessarily additive, in the sense that different positions of errors may count differently.

\subsection{Pomset metrics}\label{pomset}

Considering partially ordered multisets (pomsets), one can generalize simultaneously both the poset metrics (which essentially do not care about the alphabet structure) and the Lee metric (which is solely concerned with the cyclic structure of the alphabet). This was achieved by Sudha and Selvaraj in 2017, in \cite{Pomsets2017}. All the content of this section refer to \cite{Pomsets2017}.

Multiset is a generalization of the concept of a set that allows multiple instances of the  elements, and it had been used earlier in the context of coding theory, mainly to explore duality issues, as an alternative to the use of matroids. 

A \textbf{multiset} (or simply \textbf{mset})\index{multiset} is a pair $M=(\calx,\cf )$, where $\cf :\calx\rightarrow \N$ is the \textbf{counting function}: $\cf (i)$ counts the number of occurrences of $i$ in $M$. We define the cardinality of $M$ as $|M|=\sum_{x\in \calx}\cf (x)$ and  we denote by $M=\{k_1/a_1,\ldots ,k_n/a_n \}$ the multiset with $\cf (a_i)=k_i$. Let $\mathcal{M}^m(\calx) $ the be the family of all multisets underlying $\calx$ such that any element occurs at most $m$ times.

To define an order relation on a multiset one should consider the \textbf{Cartesian product of multisets}\index{multisets !cartesian product of} $M_1(\calx,\cf_1)$ and $M_2(\calx,\cf_2)$: it is the mset with underlying set  $\calx\times \calx$ and counting function $\cf$ defined as the product, $\cf((a,b))=\cf_1(a)\cf_2 (b)$, that is,
\[
M_1\times M_2 := \{mn/(m/a,n/b); m/a\in M_1,n/b\in M_2\}. 
\] 

A submultiset (\textbf{submset})  of $M=(\calx,\cf )$ is a multiset $S=(\calx,\cf^{\prime} )$ such that $\cf^{\prime}(x)\leq \cf (x)$, for all $x\in \calx$. We denote it by $S \ll M$. A submset $R=(M\times M,\cf_{\times})\ll M\times M$ is called a \textbf{mset relation} on $M$ if $\cf_{\times} (m/a,n/b)=mn$, for all $(m/a,n/b)\in R  $. If we consider, for example, the multiset $M=\{4/a,2/b\}$, we have that $S=\{ 5/ (4/a,2/a), 8/ (4/a,2/b) \} $ is a submset of $M\times M$, but it is not a mset relation, since $5\neq 4\cdot 2$, whereas $R= \{ 2/(2/a,1/b), 6/(2/b,3/a)  \}$ is a mset relation.
\begin{definition}
	{\em	Let $M$ be a multiset. A \textbf{partially ordered mset relation}  (or \textbf{pomset relation}) $R$ on $M$ is a mset relation satisfying:
		\begin{enumerate}
			\item $(m/a) R (m/a)$, for all $m/a\in M$ (\textbf{reflexivity});
			\item  $(m/a) R (n/b)$ and $(n/b) R (m/a)$ implies\hspace{-0.5pt} $m=n,a=b$\hspace{-1pt} (\textbf{anti-symmetry});
			\item $(m/a) R (n/b) R (k/c)$ implies $(m/a) R (k/c)$ (\textbf{transitivity}).
	\end{enumerate} }
\end{definition}

\index{pomset, partially ordered multiset} 
The pair $\mathbb{P}=(M,R)$, where $M$ is a mset and $R$ is a pomset relation is  a \textbf{partially ordered mset}  (or \textbf{pomset}). Given $\mathbb{P}=(M,R)$, a subset $I\ll M$ is called a \textbf{pomset ideal} if $m/a \in I$ and $(n/b)R(k/a)$, with $k>0$ and $b\neq a$ implies $n/b\in I$. Given a submset $S\ll M$, we denote by $\langle S \rangle_{\mathbb{P}}$ the ideal generated by $S$, that is, the smallest ideal of $\mathbb{P}$ containing $S$.


Consider the mset $M=\{r/1,r/2,\ldots ,r/n\}\in \mathcal{M}^r([n])$ where $r:=\lfloor q/2\rfloor$ is the integer part of $q/2$. The \textbf{Lee support} of a vector $\bx\in\mathbb{Z}_q^n$ is $\su_L(\bx):=\{ k/i\mid k=\pe_L (x_i)  ,k\neq 0\}$, where $ \pe_L (x_i)=\min \{x_i,q-x_i \}$ is the usual Lee weight.

The $ \mathbb{P}$\textbf{-weight} and $\mathbb{P}$\textbf{-distance} on $\mathbb{Z}_q^n$ are defined as 
\[
\pe_{\mathbb{P}}(\bx):= | \langle \su_L(\bx) \rangle_{\mathbb{P}} | \hspace{3pt}\text{ and } \hspace{3pt} d_{\mathbb{P}}(\bx,\by):=\pe_{\mathbb{P}}(\bx-\by),
\]
respectively.  \index{metric ! pomset} \index{pomset metric} 

The $\mathbb{P}$-distance satisfies the metric axioms.
In the case that $\mathbb{P}$ is an \emph{anti-chain}, that is, any two distinct pair of points $m/a,n/b\in M$ with $a\neq b$ are not comparable (neither $(m/a)R(n/b)$ nor $(n/b)R(m/a)$), we have that $\langle \su_L(\bx) \rangle_{\mathbb{P}} = \su_L(\bx)$ and so $\pe_{\mathbb{P}}(\bx)=\sum_i \pe_L(x_i) = \pe_L(\bx) $; therefore the pomset metric is a generalization of the Lee metric. In the case $M=\{1/1,1/2,\ldots ,1/n\}$ we have a poset, hence generalizing also the poset metrics.

Not much is known about pomset metrics, only what is presented in  \cite{Pomsets2017}: the authors generalize for pomsets the basic operations known for posets (direct and ordinal sum, direct and ordinal products), and study the behavior of the minimum distance under some of these operations, producing  either closed expressions or bounds.


\subsection{$m$-spotty metrics}\label{spotty}

The $m$-spotty weights were introduced by Suzuki et al. in \cite{SuzukiSpotty}, considering only the binary case. It was extended later for any finite field \cite{OzenFiniteFields} and rings \cite{sharmaSpotty}  as an extension of both the Hamming metric and the Lee metric in \cite{SiapLee}. 

Let $\bx^i=x^i_1 x^i_2 \cdots x^i_b\in \mathbb{F}_q^b$ and $\bx=\bx^1 \bx^2 \cdots \bx^n\in \mathbb{F}_q^{bn}$. We call $\bx^i$ the $i$-\emph{th $b$-byte} of $\bx$. A \textbf{spotty byte error} is defined as $t$ if  $t$ or fewer bit errors occur within a $b$-byte, for $1\leq t\leq b$.

Given $\bx=\bx^1 \bx^2 \cdots \bx^n$, $\by=\by^1 \by^2 \cdots \by^n$,  $\bx,\by\in \mathbb{F}_q^{bn}$, the  $\mathbf{(m,\ast)}$\textbf{-spotty weight} and $\mathbf{(m,\ast)}$\textbf{-spotty distance} are defined as \index{metric ! $(m,\ast)$-spotty} 
\index{$(m,\ast )$-spotty metric}
\[
\pe_{m,\ast}(\bx)=\sum_{i=1}^n\ceil*{ \frac{\pe_\ast(\bx^i)}{t}  }\text{ and } d_{m,\ast}(\bx,\by)=\pe_{m,\ast}(\bx-\by)
\]
where $\lceil x \rceil$ is the ceiling function and $\ast$ stands for any the following cases: 

\begin{itemize}
	\item The Hamming structure, with $\pe_\ast=\pe_H$;
	\item The Lee structure, with $\pe_\ast=\pe_L$;
	\item The Niederreiter-Rosenbloom chain structure, with $\pe_\ast (\bx^i)=\pe_{NR}(\bx^i)=\max\{j\mid x^i_j\neq 0 \}$. 
\end{itemize}   

Considering the $m$-spotty metrics, most of the attention was devoted to the weight distribution of a code: MacWilliams Identities were obtained for the Hamming $m$-spotty weight $\pe_{m,H}$ (\cite{SuzukiSpotty} for the binary case, \cite{OzenFiniteFields} for finite fields in general and \cite{sharmaSpotty} over rings), the Lee spotty weight $\pe_{m,L}$ in \cite{SiapLee} and the RT spotty weight $\pe_{m,RT}$ in \cite{OzenRT}.

\section{Metrics for asymmetric channels}\label{sec:asymmetric}

The \textbf{binary asymmetric channel} \index{channel ! binary asymmetric} is a memoryless channel with transition probabilities $\mathrm{Prob}(0\mid 1)=\varrho_{01}$ and $\mathrm{Prob}(1\mid0)=\varrho_{10}$, where it is assumed that $0\leq \varrho_{01}\leq \varrho_{10} \leq1/2$. The binary asymmetric channel is a model for non-volatile memories, where errors occur due to leakage of charge.   The extreme cases, $\varrho_{01}=\varrho_{10}$ and $\varrho_{01}=0$ correspond to the binary symmetric channel (see Section \ref{intro:sec:decoding}) and the $Z$-channel, respectively. 

It is possible to generalize it for a $q$-ary alphabet $\calx =\{ x_0,x_1,\ldots ,x_{q-1} \}$, by assuming that $\varrho_{ij}:=\mathrm{Prob}(x_i\mid x_j)$ is not constant for $x_i,x_j\in\calx$, $x_i\neq x_j$. Sometimes, as in \cite{Irina}, the word asymmetric is reserved for a generalization of the $Z$-channel, where $\varrho_{ij}=0$ for $i>j$.

\subsection{Asymmetric metric}
In a binary asymmetric channel we distinguish the two types  of errors: when a transmitted $0$ is received as a $1$,  it is called a $\mathbf{0}$\textbf{-error}, and when a transmitted $1$ is  received  as a $0$, it is  referred to as a  $\mathbf{1}$\textbf{-error}.
The asymmetric metric is reported in \cite{ConstantinRao} to have been  introduced in 1975 by Rao and Chawla (\cite{rao1975asymmetric}, a difficult to find reference). 
Given $\bx,\by\in\mathbb{F}_2^n$, we define $N(\bx ,\by )=|\{i\in [n] \mid x_i=0 \text{ and } y_i=1  \}$. We remark that the Hamming distance $d_H$ can be expressed as $d_H(\bx , \by )=N(\bx , \by )+N(\by , \bx )$.
The \textbf{asymmetric distance} $d_a$ between $\bx$ and $\by$ is $d_a(\bx ,\by )=\max \{N(\bx ,\by ),N(\by , \bx)  \}$. It is worth  noting that the asymmetric metric is not matched to the binary asymmetric channel, in the sense presented in Section \ref{sec:introduction}. 
\index{metric ! asymmetric}
\index{asymmetric  metric}
We say that a code $\calc\subseteq\mathbb{F}_2^n$ corrects $r$ or fewer $0$-errors if  
\[
\{ \bx\mid N(\bc_1,\bx)\leq r\}\cap \{ \bx\mid N(\bc_2,\bx)\leq r\}=\emptyset
\]
whenever $\bc_1,\bc_2\in \calc$ with $\bc_1\neq\bc_2$, and similarly for $1$-errors, exchanging $N(\bc_i,\bx)\leq r$ by $N(\bx ,\bc_i)\leq r$, $i=1,2$.

\begin{theorem}{\cite[Theorem 3]{ConstantinRao}}
	Let  $\calc\subset\mathbb{F}_2^n$ be a code with minimum distance ${d_a}(\calc)$. Then $\calc$  is capable of correcting  $r_0$  or fewer $0$-errors and $r_1$  or fewer $1$-errors, where $r_0$ and $r_1$ are fixed and $r_0+r_1<{d_a}(\calc)$. In particular, it can correct $({d_a}(\calc)-1)$  or fewer $0$-errors \textbf{or}  $({d_a}(\calc) -1)$  or fewer  $1$-errors.
\end{theorem}

If a message $\bx$ is sent and $\by$ is received, we say that the error is $t$\textbf{-symmetric} if $d_H(\bx,\by)=t$. The error is said to be $t$\textbf{-unidirectional} if $d_H(\bx,\by)=t$ and either $N(\bx,\by)=t$ or $N(\by,\bx)=t$. A code is called a $t$-\textbf{EC-AUED} code if it can correct  $t$ or fewer symmetric errors, detect $t+1$ symmetric errors and detect all ($t+2$ or more) unidirectional errors. \index{error ! $t$-unidirecional} \index{code ! $t$-EC-AUED} It is possible to prove that that a code $\calc$ is a $t$-EC-AUED code if $N(\bx,\by)\geq t+1$ and $N(\by,\bx)\geq t+1$ for all distinct $\bx,\by\in\calc$ \cite[Theorem 8]{BoseUnidirectional}. 

There are many bounds for the size of a  $t$-EC-AUED code, the lower bounds generally presented as parameters of specific codes. Many of these bounds can be found in a comprehensive survey, with bibliography updated in 1995, due to  Kl\o ve \cite{Klove}. We present just two of the most classical upper bounds, whose proofs can also be found in \cite{Klove}.\index{bound ! Lin and Bose ! for asymmetric metric}\index{bound ! Varshamov ! for asymmetric metric}  
\index{Lin and Bose  bound for asymmetric metric} \index{Varshamov  bound for asymmetric metric}
\begin{theorem}
	Let $A(n,t)$ denote the maximal size of a $t$-EC-AUED code of length $n$. The following bounds hold:
	\begin{eqnarray*}
		\text{ \emph{Lin and Bose  \cite[Theorem 2.5]{linbose}}}: & A(n,1) \leq \dfrac{2}{n}\dbinom{n}{\lfloor n/2 \rfloor} \\
		\text{\emph{Varshamov  \cite{Varshamov}}}: & A(n,t) \leq  \frac{\displaystyle 2^{n+1}}{\displaystyle \sum_{j=1}^{t} \bigg\{  \binom{\lfloor n/2 \rfloor}{j} +\binom{\lceil n/2 \rceil}{j}  \bigg\}  }.
	\end{eqnarray*} 
\end{theorem}

\subsection{Generalized asymmetric metric}

The asymmetric metric, originally defined over a binary alphabet, can be generalized over a $q$-ary alphabet $\llbracket q \rrbracket := \{0,1,\ldots,q-1  \}$, or the alphabet $\mathbb{Z}_\geq$, consisting of nonnegative integers. 
\index{metric ! generalized asymmetric} \index{generalized asymmetric metric}

Given $\bx,\by\in(\mathbb{Z}_\geq )^n$, we define the \textbf{generalized asymmetric weight}  and the  \textbf{generalized asymmetric metric}  respectively as 
\[
N_g(\bx,\by):=\sum_{i=1}^n\max\{x_i-y_i,0 \} \text{ and } d_{g}(\bx,\by)=\max\{N(\bx,\by),N(\by,\bx)\}.
\]
This distance was recently introduced  in \cite{DNA}, where it is used to define a metric on the so called $l$-\emph{gramm profile} of $\bx\in \llbracket q \rrbracket ^n$. It turns out it is an interesting distance for the types of errors that may occur in the DNA storage channel, which includes substitution errors (in the synthesis and sequencing processes) and coverage errors (which may happen during the DNA fragmentation process). For details see \cite{DNA}.

\section{Editing metrics}\label{sec:editing}

An \textbf{editing distance} is a measure of similarity between strings (not necessarily of the same length) based on the minimum number of operations required to transform one into the other. Different types of operations lead to different metrics, but the most common ones are insertions, deletions, substitution and transpositions.\index{metric ! editing } \index{editing metric}

Consider an alphabet $\calx$ and let $\calx^{\ast} $ be the space of all finite sequences (strings) in $\calx$.  Given two strings $\bx=x_1x_2\cdots x_m$ and $\by=y_1y_2\cdots y_n$, there are four basic errors, described as operations from which $\by$ can be obtained from $\bx$.
\begin{enumerate}
	\item \textbf{(I) Insertion}: In case $\by$ is obtained by inserting a single letter to $\bx$ ($n=m+1$);
	\item \textbf{(D) Deletion}: In case $\by$ is obtained by deleting a single letter of $\bx$ ($n=m-1$);
	\item \textbf{(S) Substitution}: In case $\by$ is obtained by substituting a single letter of $\bx$ ($n=m$);
	\item \textbf{(T) Transposition}: In case $\by$ is obtained by transposing two adjacent letters of $\bx$; that is, for some $i$ we have $x_i=y_{i+1},x_{i+1}=y_i$ ($n=m$).
\end{enumerate}

These operations are considered to be an appropriate measure to describe human typeset misspellings. In\cite{damerau}, Damerau claims that 80 percent of the misspelling errors have distance one in the metric admitting all the four types of basic errors. It is also has applications in genomics, since DNA duplication is  commonly  disturbed by the considered operations, each operation occurring with similar probabilities.

\begin{definition}
	{\em 	Given a set $\cale$ of operations, $\{I,D\}\subseteq \cale\subseteq \{I,D,S,T \}$, the distance $d_\cale$ between strings $\bx$ and $\by$ is the minimal number of operations in $\cale$ needed to get $\by$ from $\bx$. For simplicity, we will write $d_{ID}$ instead of $d_{\{I,D\}}$, and similarly to other subsets $\cale$. }
\end{definition}
For $\cale =\{I,D\}$, we have \textbf{Levenshtein's insertion-deletion metric}, introduced in \cite{Levens-Inser-Delet-English} for the case of binary alphabets. The case $\cale =\{I,D,S\}$ is known as \textbf{Levenshtein's editing metric}, also introduced in \cite{Levens-Inser-Delet-English} for the binary case.  In this work, Levenshtein proved \cite[Lemma 1]{Levens-Inser-Delet-English} that a code capable of correcting $t$ deletions and separately correcting $t$ insertions can correct a mixture of  $t$ insertions and deletions. \index{metric ! Levenshtein's insertion-deletion}\index{metric ! Levenshtein's editing}
\index{Levenshtein's insertion-deletion metric}\index{Levenshtein's editing metric}

Denoting by $|\bx |$ the length of the string, in \cite{Hollmann} it was proved that $d_{ID}(\bx,\by)=|\bx |+|\by |-2\rho (\bx,\by)$, where $\rho (\bx,\by)$ is the maximum length of a common subsequence of $\bx$ and $\by$. As observed in \cite{Hollmann}, it is important to distinguish between the minimum number of insertions/deletions and minimum number of insertions and deletions. If one defines $d_{ID}^\ast(\bx,\by)=2e$ meaning that $e$ is the smallest number in which $\bx$ can be transformed into $\by$ by at most $e$ insertions and at most $e$ deletions, we get a distinct metric, with $d_{ID}(\bx,\by)\leq d_{ID}^\ast (\bx,\by)$.

\begin{theorem}{\cite[Theorems 4.1 and 4.2]{Hollmann}} A code $\calc\subseteq \calx^\ast$ is capable of correcting $e$ insertions/deletions if and only if $d_{ID}(\calc)> 2e$ and it is capable of correcting $i$ insertions and $d$ deletions if and only if  $d_{ID}^\ast(\calc)>2(d+i)$.
\end{theorem} 

The structure of the edit space, that is, the structure of  $\calx^{\ast} $ equipped with the metric $d_{IDS}$, is studied in \cite{Houghten}. The group $GL(\calx^{\ast},d_{IDS})$ is described as the product $GL(\calx^{\ast},d_{IDS})=\il \gamma \ir \times S_q$ where $\gamma:\calx^\ast \rightarrow \calx^\ast$ is the reversion map $\gamma (x_1x_2\cdots x_m)=x_mx_{m-1}\cdots x_1$, $\il \gamma \ir$ is the group generated by $\gamma$ and $S_q$ is the group of permutations of the alphabet $\calx$ ($|\calx |=q$), acting as usual on every position of a string, $\sigma (x_1x_2\cdots x_m)=x_{\sigma(1)}x_{\sigma(2)}\cdots x_{\sigma(m)}$.

\bigskip

\subsection{Bounds for editing codes}

Denote by $A_{\cale}(n,t)_q$ the maximal size of a $q$-ary code of length $n$ capable of correcting $t$ errors of the types belonging to $\cale$. The asymptotic behavior of $A_{\cale}(n,t)_q$ was first studied by Levenshtein in \cite{Levens-Inser-Delet-English}, showing that  $A_{ID}(n,1)_2$  behaves  asymptotically as $2^n/n$. 
Lower and upper bounds for the $q$-ary case were given in \cite{Levenshteinqary}:\index{bound  ! for editing metrics}
\[
\frac{q^{n+t}}{\left( \sum_{i=0}^t\binom{n}{i}(q-1)^i \right)^2}\leq A_{ID}(n,t)_q\leq \frac{q^{n-t}}{\sum_{i=0}^t\binom{n-t}{i}}+\sum_{i=0}^{n-2}\binom{n-1}{i}(q-1)^i.
\]
It was proved in \cite[Theorem 2]{Cullina} that a code capable of correcting $a$ deletions and $b$ insertions has size asymptotically bounded above by \[\frac{q^{n+b}}{(q-1)^{a+b}\binom{n}{a+b}\binom{a+b}{b}} .\]
\medskip

The case $\cale =\{I,D,S\}$ is also approached in \cite{Levens-Inser-Delet-English} where Levenshtein  gives the bounds for codes capable of correcting one insertion-deletion-substitution error:
\[
\frac{2^{n-1}}{n}\leq A_{IDS}(n,1)_q\leq\frac{2^n}{n+1}.
\]
As a general relation, useful for studying bounds, we can find in \cite{Houghten}: 
\[A_{IDS}(n,t)_q\leq q\cdot A_{IDS}(n-1,t)_q.\]
Tables  of known values of  $A_{IDS}(n,t)_q$ and  $A_{ID}(n,t)_q$ (for small values of $n,t$) can be found in \cite{table1} for $q=2$ and \cite{table2} for $q=3$.

%

\section{Permutation metrics}\label{sec:permutation}

The symmetric group $S_n$ acts on $\calx^n$ permuting the coordinates: \[\sigma (\bx)=x_{\sigma(1)} x_{\sigma(2)} \cdots x_{\sigma(n)}  ,\] for $\sigma\in S_n,\bx\in\calx^n$. Considering the group structure of  $S_n$ or a subgroup $G\subset S_n$ is a common procedure to handle algorithm difficulties. It is used, for example, when considering a code to be an orbit (best if with no fixed-points) of such a group.    Considering  permutations as codewords is what is known as \textbf{code in permutations}, and it is relevant in coding for flash memories and to handle problems concerning rankings comparisons.  It was in the context of ranking comparisons that, in 1938, Kendall introduced  \cite{Kendall} the metric which became known as the {Kendall-tau metric}.
\medskip

We denote a permutation $\sigma\in S_n$ as $[\sigma(1),\sigma(2),\ldots ,\sigma(n)  ]$. The (group) product  of permutations $\sigma ,\pi\in S_n$ is defined by  $\pi\circ\sigma (i)=\sigma (\pi (i))$, for $i\in [n]$.  A transposition $\sigma :=(i,k)$ is a permutation  such that 
\[\sigma (i)=k,\sigma(k)=i \text{ and } \sigma(j)=j, \text{ for } j\neq i,j.\]
We say the transposition is adjacent if $k=i+1$. The \textbf{Kendall-tau metric} $d_\tau(\sigma, \pi)$ is the minimal number of adjacent transpositions needed to transform $\sigma$ into $\pi$. The formula
\[
d_\tau(\sigma,\pi)= | \{ (i,j)\mid \sigma^{-1}(i)<\sigma^{-1}(j) \text{ and } \pi^{-1}(i)>\pi^{-1}(j)  \}  |,
\]
is well known, and is  actually the original formulation of Kendall.\index{metric ! Kendall-tau}\index{Kendall-tau metric}

We denote by $A_{d_\tau}(n,t)$ the maximum size of a code $\calc\subseteq S_n$ with minimum distance ${d_\tau}(\calc)=t$. In \cite{BargMazumdar} we find various bounds for $A_{d_\tau}(n,t)$, among others the Singleton and Sphere Packing Bounds.\index{bound ! Singleton ! for Kendall-tau metric}\index{bound ! Sphere Packing ! for Kendall-tau metric} 
\index{Singleton bound for Kendall-tau metric} \index{Sphere Packing bound for Kendall-tau metric} 

\begin{theorem}{\cite[Theorems 3.2]{BargMazumdar}}
	\begin{enumerate}
		\item For $n-1<t<n(n-1)/2$,  \[ A_{d_\tau}(n,t)\leq \lfloor  3/2 +\sqrt{n(n-1)-2t+1/4}\rfloor ! . \]
		\item \[  \frac{n!}{|B_{d_\tau}(2r)|}\leq A_{d_\tau}(n,2r+1)\leq \frac{n!}{|B_{d_\tau}(r)|},
		\]
		where $B_{d_\tau}(R)$ is the $d_\tau$-ball of radius $R$.
	\end{enumerate}
\end{theorem}

Obstruction conditions for the existence of perfect codes with the Kendall-tau metric can be found in \cite{Buzaglo}.

It is worth remarking that the set $S=\{(1,2),(2,3),\ldots ,(n-1,n)  \}$
of adjacent transpositions is a  set of generators of $S_n$, minimal if considering only transpositions as generators. Moreover, a transposition $\sigma=(i,j)$ is its own inverse ($\sigma =\sigma^{-1}$). It follows that the Kendall-tau metric is just the graph metric in the Cayley graph of $S_n$ determined by the generator set $S$ (the vertices are the elements of the group and two vertices are connected by an edge if and only if they differ by a generator). This is actually the Coxeter group structure. The group $S_n$ acts with no fixed points on the subspace $V=\{\bx\in\mathbb{R}^n\mid x_1+x_2+\cdots +x_n=0   \}$ and there is a simplicial structure on the unit sphere of $V$ (known as the Coxeter complex) on which $S_n$ acts simply transitively. This action makes Coxeter groups suitable for spherical coding; see Definition \ref{SphericalCode} for the definition of a spherical code. The study of such group-codes, a generalization of Slepian's permutation modulation codes, is studied in \cite{Mittel}.

\medskip
There are other relevant metrics on $S_n$. Let $G_n:=\mathbb{Z}_2\times \mathbb{Z}_3\times \cdots \times \mathbb{Z}_n$ and consider the embedding $:S_n \rightarrow G_n$ which associates to $\sigma\in S_n$ the element $\bx_\sigma=x_\sigma (1) x_\sigma (2) \cdots x_\sigma(n-1)\in G_n$, where $x_\sigma $ is defined by 
\[
x_\sigma (i)=|\{j\mid j<i+1,\sigma(j)>\sigma (i+1)  \} |,\] 
for $i=1,2,\ldots ,n-1$.
The $\ell_1$ metric on $G_n$, $d(\bx,\by)=\sum_{i=1}^{n-1}|x(i)-y(i)|$, induces a metric on $S_n$: $d_{\ell_1}(\sigma,\pi):=d_{\ell_1}(\bx_\sigma ,\bx_\pi)$. This is a metric related to the Kendall-tau metric by the relation $d_\tau(\sigma,\pi)\geq d_{\ell_i}(\sigma,\pi)$. More on permutation metrics can be found in \cite{BargMazumdar}.

%
%



%



%
\bibliography{bib_ency_A}
\bibliographystyle{plain}
%

\end{document}